# Data Synthesis based on Generative Adversarial Networks

Noseong Park*
University of North Carolina
Charlotte, NC, USA
npark2@uncc.edu

Mahmoud Mohammadi*
University of North Carolina
Charlotte, NC, USA
mmoham12@uncc.edu

Kshitij Gorde
University of North Carolina
Charlotte, NC, USA
kgorde@uncc.edu

Sushil Jajodia
George Mason University
Fairfax, VA, USA
jajodia@gmu.edu

Hongkyu Park
ETRI
Daejeon, South Korea
hkpark@etri.re.kr

Youngmin Kim
ETRI
Daejeon, South Korea
injesus@etri.re.kr

## ABSTRACT

Privacy is an important concern for our society where sharing data with partners or releasing data to the public is a frequent occurrence. Some of the techniques that are being used to achieve privacy are to remove identifiers, alter quasi-identifiers, and perturb values. Unfortunately, these approaches suffer from two limitations. First, it has been shown that private information can still be leaked if attackers possess some background knowledge or other information sources. Second, they do not take into account the adverse impact these methods will have on the utility of the released data. In this paper, we propose a method that meets both requirements. Our method, called *table-GAN*, uses generative adversarial networks (GANs) to synthesize fake tables that are statistically similar to the original table yet do not incur information leakage. We show that the machine learning models trained using our synthetic tables exhibit performance that is similar to that of models trained using the original table for unknown testing cases. We call this property *model compatibility*. We believe that anonymization/perturbation/synthesis methods without model compatibility are of little value. We used four real-world datasets from four different domains for our experiments and conducted in-depth comparisons with state-of-the-art anonymization, perturbation, and generation techniques. Throughout our experiments, only our method consistently shows balance between privacy level and model compatibility.

## 1. INTRODUCTION

In the big data era, sharing data with partners or releasing data to the public frequently occurs. Privacy should be the top priority in the process to protect people who were willing to share information.

Anonymization techniques remove identifiers (such as social security numbers) and modify quasi-identifiers (such as gender, ZIP code, age, occupation, and so forth). However, other sensitive attributes that are neither identifiers nor quasi-identifiers are often disclosed without any modification. If adversaries possess background knowledge or other information sources, then they can recover the identification of records (i.e., *re-identification attack*). Data perturbation is occasionally preferred because it changes or adds noise to values. Unfortunately, data usability is negatively impacted after these modifications. Moreover, methods have been developed to recognize and remove noise [10].

In fact, it is very difficult to simultaneously achieve good privacy and usability levels after anonymization or other modifications. In general, privacy level and data utility are inversely proportional to each other. We propose a data synthesis method based on generative adversarial networks (GANs). GANs are a generative model very recently proposed by deep learning researchers [19]. These models have shown significant improvements over other generative models in image and text datasets. Our method, named *table-GAN*, is specialized for synthesizing tables that contain categorical, discrete, and continuous values — we leave other types of data as future work. The main advantages of generating synthetic tables are as follows:

- There is no one-to-one relationship[1] between real records and synthetic records, and re-identification attacks are impossible.

- All attribute values are fake and safe from attribute disclosure.

- Machine learning models trained using very carefully synthesized tables show behavior similar to that of models trained using the original table; they can replace each other (i.e., model compatibility).

- Synthesized tables can be shared with partners without any concerns of information leakage.

However, there is one critical point in our deep learning-based approach. Membership attack is a recently proposed concept to infer about training samples by observing outputs from machine learning models [33]. Even though re-identification attack and attribute disclosure are prevented, it is useless if membership attacks can be done with high accuracy. Therefore, we show that our table-GAN is strong against all those attacks: re-identification attack, attribute disclosure, and membership attack.

---

*Equally contributed

[1] An anonymized (or perturbed) table is created by modifying records in the original table one by one and there typically exists one-to-one correspondence between the two tables, which is the main reason why re-identification attacks are possible.

Whereas original GANs consist of two neural networks (*generator* and *discriminator* neural networks), our table-GAN consists of three neural networks (*generator*, *discriminator*, and *classifier* neural networks). The discriminator attempts to distinguish between real and synthetic records (i.e., binary classification), and the generator obfuscates the task of the discriminator by generating realistic records. They continue iterating the adversarial game, and the generator can achieve unprecedented generation performance at the end of the two-player game. In our table-GAN, we add an additional classifier neural network to increase the **semantic integrity** of synthetic records. For instance, (cholesterol=60.1, diabetes=1) is not a semantically correct record (because the cholesterol level is too low to be diagnosed as diabetes), and there may be no such record in the original table. We prevent the generation of such records by adding a classifier (that learns the semantics from the original table) into the training process because otherwise it is easy to determine that the table is fabricated.

Loss (or objective) functions are key for training neural networks. In general, neural networks are trained by minimizing loss functions. For instance, Equation (1) shows the objective function of conventional GANs, denoted as *original loss* in our paper. In addition to this function, we also design two additional loss functions – *information loss* and *classification loss* – that are specialized in the table synthesis process.

*Information loss* matches the first-order (i.e., mean) and second-order (i.e., standard deviation) statistics of record features; thus, synthetic records have the same statistical characteristics as the original records. We use the concept of maximum-margin in the hinge loss [31] to control the quality of synthesis process. *Classification loss* maintains the semantic integrity. We found that synthesizing a semantically sound table while maintaining a good balance between privacy and usability is very challenging. Therefore, our training process is considerably more complicated than the original GAN model; however, in our experiments, the training time is less than 20 minutes.

For our experiments, we use four datasets from different domains and consider many state-of-the-art anonymization, perturbation, and synthesis techniques. Among all methods, the proposed table-GAN shows the best trade-off between privacy and model compatibility. We also applied the state-of-the-art membership attack method [33] to attack our table-GAN and its result shows that our hinge loss-based privacy control mechanism can effectively prevent it.

## 2. RELATED WORK

We performed an extensive literature survey, and we introduce important works in this section. First, we define the following key terms:

- A *table* means a relation or table of a relational database. It consists of attributes (i.e., columns) and records (i.e., rows).

- An *identifier* is an attribute that assigns a unique number to each record, such as social security number (SSN).

- A *quasi-identifier* (QID) is not a unique identifier; however, a combination of QIDs is occasionally sufficient to identify a record, such as occupation, age, and ZIP code.

Table 1: Original Table. ZIP and Age are QIDs; Salary and Disease are sensitive attributes.

| No | ZIP | Age | Salary | Disease |
|---|---|---|---|---|
| 1 | 47677 | 29 | 3K | AIDS |
| 2 | 47672 | 22 | 4K | Ebola |
| 3 | 47678 | 27 | 5K | Cancer |
| 4 | 47905 | 53 | 6K | AIDS |
| 5 | 47909 | 52 | 11K | Ebola |
| 6 | 47906 | 57 | 8K | Heart Disease |

Table 2: Table anonymized by 3-anonymity and 3-diversity. There are two equivalence classes and in each equivalence class, there are three records that are not distinguishable w.r.t. QIDs (i.e., 3-anonymity) and their sensitive values are all different (i.e., 3-diversity).

| No | ZIP | Age | Salary | Disease |
|---|---|---|---|---|
| 1 | 4767* | $\leq 40$ | 3K | AIDS |
| 2 | 4767* | $\leq 40$ | 4K | Ebola |
| 3 | 4767* | $\leq 40$ | 5k | Cancer |
| 4 | 4790* | $\geq 50$ | 6K | AIDS |
| 5 | 4790* | $\geq 50$ | 11K | Ebola |
| 6 | 4790* | $\geq 50$ | 8K | Heart Disease |

- *Sensitive attribute (information)* generally means all other attributes except for identifiers and QIDs, such as grade point average (GPA), salary, disease status, and so on.

### 2.1 Privacy Preserving Method

To protect against re-identification attacks, various privacy models have been introduced. The re-identification attack for an individual (or a group of people) can be conducted by linking some set of attributes in the published dataset with an external dataset to identify the target individual or groups. This set of attributes, such as ZIP code, birthday, gender, and so on, are called QIDs. The goal of many privacy preserving techniques is to transform QIDs in such a way that they cannot be linked together to identify a particular person.

One of the most fundamental and widely adopted privacy models is $k$-*anonymity* introduced by Samarati and Sweeney [32]. This model introduced the concept of the *equivalence class* of records, where one record is similar to at least $k-1$ other records in the same equivalence class with respect to their QIDs. In other words, it modifies QIDs, and records with the same modified QIDs constitute an equivalence class (see Tables 1 and 2).

As we describe below, there exist other types of attacks, and other notions have been proposed to mitigate them. Nonetheless, $k$-anonymity is still used in the healthcare world, in large part because of its simplicity and utility preservation compared to other definitions[2].

Although $k$-anonymity reduces the possibility of re-identification attacks, adversaries can still obtain information about other sensitive attributes of that table (because existing methods focus on modifying QIDs after leaving other sensitive information unaltered). This sensitive information enables adversaries to conduct homogeneity and background knowledge attacks [24], also known as *attribute disclosure*. To protect against these attacks, the authors of [24] introduced *l-diversity* to ensure that the sensitive attributes of each equivalence class have at least $l$ different values. The $l$-diversity is effective in protecting categorical attributes (because continuous attributes with $l$ diverse values are not sufficient) but is still vulnerable in cases where the adversaries know the global distributions of sensitive attributes.

---
[2]See, for example, the webite https://desfontain.es/privacy/k-anonymity.html

Consequently, the authors of [22] introduced $t$-*closeness* to ensure that the distributions of sensitive attributes in each equivalence class are similar to their global distributions. $\delta$-disclosure is also a concept to protect sensitive attributes from re-identification attacks [14]. Note that $t$-closeness and $\delta$-disclosure do not change sensitive values but construct equivalence classes in a way that reduces the possibility of re-identification attacks.

Perturbation is also very popular for statistical disclosure control (SDC) [18]. Adding additive or multiplicative noise to continuous values is one of the most popular perturbation techniques. However, it is also very popular to study the removal of noise and recovery of the original data in many related fields [9]. Thus, other perturbation techniques, such as micro-aggregation and post-randomization method (PRAM), have been also developed, and they can perturb continuous and categorical values, respectively. In particular, PRAM mainly aims at modifying sensitive attributes.

Existing data anonymization/perturbation methods provide reasonable model compatibility in many cases because they do not actively modify sensitive attributes as in Tables 1 and 2. However, there also exits a non-trivial possibility of information leakage. In general, their balance between privacy level and model compatibility is not satisfactory — we will show this in our experiments.

One more related privacy concept is $\epsilon$-differential private data release [26]. $\epsilon$-differentially private data is created by drawing perturbed samples (more precisely, $\epsilon$-differential samples) from the original dataset. In general, data utility is significantly decreased after this process.

A few researchers have focused on generating synthetic data, e.g., the condensation method [8]. It first finds clusters of records that share similar values and synthesizes them on top of several statistical assumptions for technical convenience and without any attention to the semantic integrity. As the clustering problem is NP-hard and even its approximation takes non-trivial time, their scalability is not satisfactory. Our deep learning-based method can generate semantically correct records after learning any complicated table without relying on any statistical assumption about data. Because it is very easy to detect that a table is synthesized after identifying semantically incorrect records, our contributions are significant.

## 2.2 Risk Evaluation Methods

Developing privacy risk evaluation methods is also an independent research topic. However, these methods are all designed for anonymization and perturbation. Three popular risk evaluation metrics are based on the prosecutor, journalist, and marketer attacker models [17]. They measure the percentage of re-identified records given a certain attacker model. In the prosecutor model, the attacker already knows about QIDs of all target people and tries to uncover their sensitive attributes. Thus, the successful re-identification probability for a certain person $p$ is simply calculated as

$$risk(p) = \frac{1}{\text{the size of the matching equivalence class to } p}.$$

In the journalist and marketer models, the attacker does not have specific targets but does his/her best to re-identify as many records as possible using available background information. In general, these two models are weaker than the prosecutor model in several points and they also require equivalence classes to calculate risk

---

**Input:** Real Samples: $\{x_1, x_2, \cdots\} \sim p(x)$
**Output:** a Generative Model $G$
1   $G \leftarrow$ a generative neural network
2   $D \leftarrow$ a discriminator neural network
3   **while** *until convergence of loss values* **do**
4      Create a mini-batch of real samples $X = \{x_1, \cdots, x_n\}$
5      Create a set of latent vector inputs $Z = \{z_1, \cdots, z_n\}$
6      Train the discriminator $D$ by maximizing Equation (1)
7      Train the generator $G$ by minimizing Equation (1);
8   **end**
9   **return** $G$

**Algorithm 1:** Training algorithm of GANs

scores. In our method, we do not create any equivalence class but disclose full synthetic values. Therefore, this risk evaluation cannot be applied.

The authors of [34] proposed a risk evaluation method based on entropy. Its formula requires the average number of correct recalls in the one-to-one correspondence between the original and anonymized tables. This cannot be measured for our method.

## 2.3 Generative Adversarial Network

Generative adversarial networks (GANs) are a recently developed generative model [19] to produce synthetic images or texts after being trained. The learning process in the model is based on one generator ($G$) and one discriminator ($D$) neural networks playing the following zero-sum minimax (i.e., adversarial) game:

$$\min_G \max_D V(G,D) = \mathbb{E}[\log D(x)]_{x \sim p_{data}(x)} + \mathbb{E}[\log(1 - D(G(z)))]_{z \sim p(z)}, \quad (1)$$

where $p(z)$ is a prior distribution of latent vector $z$, $G(\cdot)$ is a generator function, and $D(\cdot)$ is a discriminator function whose output spans $[0,1]$. $D(x) = 0$ (resp. $D(x) = 1$) indicates that the discriminator $D$ classifies a sample $x$ as *generated* (resp. *real*).

Algorithm 1 shows the general training concept of GANs. $G$ and $D$ can be any form of neural networks. The discriminator $D$ attempts to maximize the objective, whereas the generator $G$ attempts to minimize the objective. In other words, the discriminator $D$ attempts to distinguish between real and generated samples, while the generator $G$ attempts to generate realistic fake samples that the discriminator $D$ cannot distinguish from real samples. One can also consider the discriminator as a teacher and the generator as a student. The teacher provides feedback to the student on the quality of work.

Although there are many variations, we design our table-GAN methods on DCGAN [30] because it is considered to be the most mature model [12], and many other GANs rely on its neural network architectures.

Two recent studies applied GANs based on *recurrent neural network (RNN) architectures* to synthesize only discrete values in electronic health records (EHR) [15] and patent records [16]. One other research also applied the same RNN techniques to synthesize time-series values [11]. Our table-GAN aims at synthesizing general relational tables that consists of various data types based on *convolutional neural network (CNN) architectures*. RNNs (resp. CNNs) are widely used for natural language processing (resp. computer vision). There is one famous example describing their difference. In computer vision, Red $-\tau$ = Pink[3], where $\tau$ is a small number, is semantically valid (i.e., continuous data type) but in natural language processing, Penguin $-\tau$ = Ostrich cannot be defined (i.e., discrete data type). Thus, [15] cannot synthesize general relational databases. However, our method can generate both continuous and discrete values after some tricks.

---
[3]Recall that colors are basically numbers in the RGB code space.

## 2.4 Membership Attack

The membership attack is a recently introduced concept that can be applied to general machine learning algorithms [33]. Its goal is to infer about training samples after observing outputs of a target machine learning algorithm. The attack model assumes that attackers have black-box access to the target model to attack and know its detailed algorithm design and architecture. Utilizing the black-box access, attackers can create training samples for their attack models. In their paper, however, they presented a method to attack only classification algorithms and it requires more studies to attack other types of machine learning models. We customize their attack method to attack our table-GAN.

## 3. OVERALL ARCHITECTURE

We introduce security and privacy concerns that we will address in this paper and the overall workflow of the proposed table-GAN.

### 3.1 Security & Privacy Concerns to address

We address the following three privacy and security concerns in this paper: re-identification attack, attribute disclosure, and membership attack. Since table-GAN generates fully synthetic tables, it is strong against re-identification and attribute disclosure issues by nature. The entire table is completely synthesized in our method so no real records are directly disclosed and attackers cannot reveal their original identifications.

To prevent membership attacks, we use the hinge loss for training the generator (see Equation (4)). If the quality of synthesized records are too high, attackers can easily infer about its original table. The role of the hinge loss is to slightly disturb the training process of table-GAN so that it converges to a point that balances the synthesis quality and the possibility of being attacked.

### 3.2 Overall Workflow

The overall workflow of our approach is presented in Figure 1. It is processed in the following sequence:

1. Records in the original table are converted into square matrix form; if needed, we pad with zeros. For example, a record that consists of 24 values can be converted into a $5 \times 5$ square matrix after padding a zero. Other option is to input records in the original vector format and perform 1D convolutions. However, its synthesis performance is sub-optimal due to its limited convolution computations, compared to the proposed strategy, in our preliminary experiments.
2. The proposed table-GAN is trained using the converted square matrices — we will describe the details of the table-GAN in Section 4.
3. The table-GAN generates many synthetic square matrices (i.e., synthetic records) that will be converted and merged into a table.
4. The generated fake table is shared with partners who will perform analyses and design machine learning models.
5. The machine learning model trained using the synthetic table should be able to replace the model trained using the original table. In particular, we call this property *model compatibility*.

The generation process should have parameters to trade off between the level of privacy and model compatibility. By decreasing the level of privacy, we can make synthetic records similar to the ones in the original table and improve model compatibility.

## 4. PROPOSED METHOD

We introduce the proposed table-GAN that, given a table to synthesize and several parameters to control the privacy level, generates a synthetic table that is statistically similar to the original table.

### 4.1 Neural Network Architecture

The deep convolutional GAN (DCGAN) is one of the most influential works for GANs. Many meaningful ideas have been proposed to design the neural network architecture of the generator ($G$) and the discriminator ($D$), as follows: (1) replacing spatial pooling functions with strided convolutions, (2) eliminating fully connected hidden layers, and (3) using batch normalization [20], ReLU for the generator [27], and LeakyReLU [23] for the discriminator.

Our model also adopts the DCGAN architecture but has an additional neural network called *classifier* ($C$), as follows:

- A generator neural network $G$ to produce synthetic records that have the same distribution as that of real records;

- A discriminator neural network $D$ to distinguish between real and synthetic records;

- A classifier neural network $C$ to predict synthetic records' labels. We found that adding $C$ helps maintain the consistency of values in the generated records. For instance, a record with gender = "Male" and disease = "Uterine Cancer" can be prevented; the classifier learns about the consistency from the original table. There exist some other GAN models that have the same approach to use axillary classifiers [28]. They also showed that adding more classifiers can improve the generation quality significantly.

We describe neural network architectures in more detail in the following.

#### 4.1.1 Discriminator

The discriminator network $D$ is a neural network trained to classify the generated records as *synthetic* and the records in the original table as *real*. Precisely, $D$ is a convolutional neural network (CNN) that contains multiple layers. In each layer, a list of learnable filters (e.g., 3×3 matrices) are applied to the entire input matrix (i.e., convolution operations). Recall that records are converted into square matrices in our method. Thus, the layer output size is proportional to the number of filters in each layer.

The output of a layer is the input to the next layer, as shown in Figure 2. The dimension of intermediate tensors keeps decreasing and the depth continues increasing until the last sigmoid activation layer that generates the probability of being real or synthetic. There are other intermediate layers (omitted in Figure 2) that affect the functionality of the network, such as batch normalization and LeakyReLU [30].

The input to the first layer is a $d \times d$ matrix that represents one real or synthetic record. The discriminator is trained to predict 1 for real records and 0 for synthetic records after the last sigmoid layer.

If the number of attributes in the original table is less than the input size, then each record is padded with zeros and reshaped into a square matrix. Our model is configurable and able to learn from a table with many attributes (e.g., $16 \times 16 = 256$ attributes in a single table and some GANs supports images that are as large as $1024 \times 2014$ [21]).

#### 4.1.2 Generator

The generator $G$ is also a neural network that consists of multiple de-convolutional layers. As shown in Figure 2, $G$ performs a process that is opposite to that of the discriminator. Its input is a latent vector $z$ that is uniformly sampled from the unit hypercube space[4]. Multiple de-convolutional layers convert the input $z$ to a 2-dimensional matrix corresponding to a record in the synthetic table.

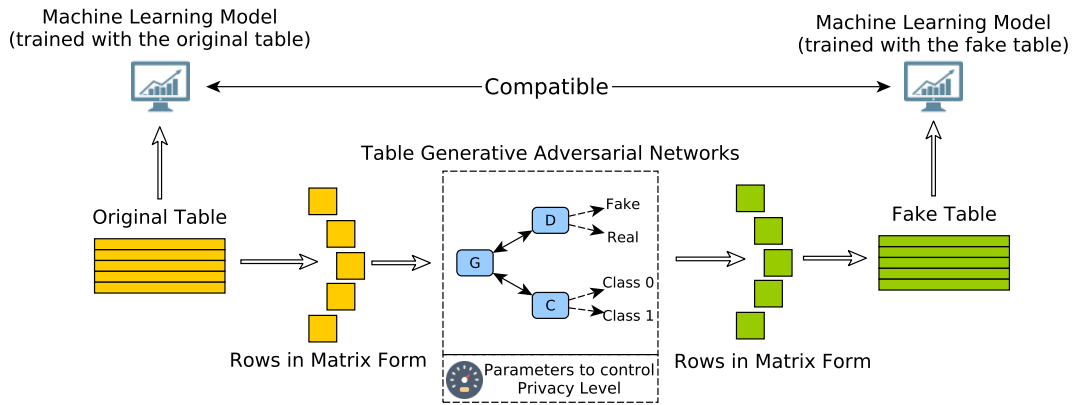

Figure 1: The overall workflow of the proposed method. The fake table (marked in green) is generated by the proposed table-GAN trained using the original table (marked in yellow). Machine learning models trained using the fake table should show the same behaviors as models trained using the original table (i.e., model compatibility). Our goal is to achieve general model compatibility regardless of machine learning algorithms and tasks.

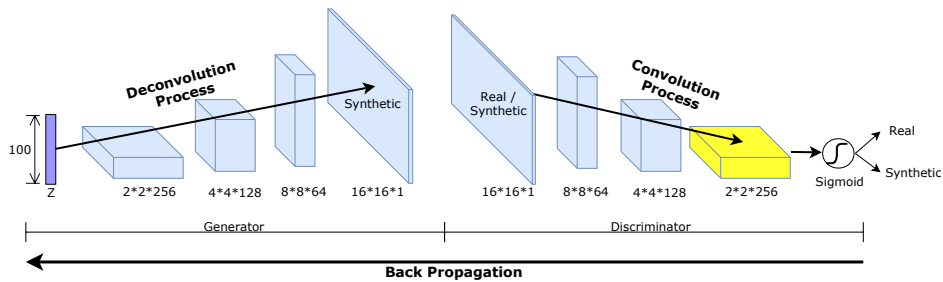

Figure 2: Our table-GAN architecture. The classifier is omitted because of space limitations, and it has the same neural network architecture as the discriminator. The generator (resp. discriminator) performs a series of deconvolution (resp. convolution) operations to generate (resp. classify) a record. The final loss after the sigmoid activation can be back-propagated to the generator. The dimensions of the latent vector input $z$ and intermediate tensors should be configured considering the number of attributes (e.g., $16 \times 16 = 196$ attributes in this figure).

The generator can be trained by the discriminator's prediction result (in fact, this is one of the multiple ways we propose to train the generator in our table-GAN). Its goal is to deceive the discriminator. This training process can be very efficiently implemented by back-propagation.

### 4.1.3 Classifier

The classifier network $C$ has the same neural network architecture as the discriminator. However, it is trained by ground-truth labels in the original table. Therefore, the classifier is trained to learn about the correlation between labels and other attributes from the table. Given a synthetic record, it can teach the generator whether the record is semantically correct. For example, (cholesterol=50, diabetes=1) is not a correct record because cholesterol=50 is too low to be diagnosed as diabetes. If such records are synthesized by the generator, the classifier can effectively correct the generator.

In fact, the discriminator itself can learn about the semantic integrity to some degree. Semantically incorrect generations are likely not to be classified as real by the discriminator. However, it does not mainly consider about the semantic integrity and we found some incorrect generation examples without the classifier.

[4]A high-dimensional space where each dimension is confined to the range of [-1,1]

### 4.2 Loss Functions

The loss function contains the philosophy for training neural networks. In general, neural networks are trained by minimizing a loss function, and ill-designed loss functions deteriorate the training process and lead to malfunctioning neural networks. Therefore, the design of loss functions to train the three proposed neural networks is key for successful table syntheses. We adopt one loss function from DCGAN and design two more loss functions, as follows:

- The *original loss* is adopted from DCGAN and was already shown in Equation 1.

- The *information loss* is defined as the discrepancy between two statistics of synthetic and real records.

- The *classification loss* is defined as the discrepancy between the predicted label by the classifier and the synthesized label.

The discriminator is trained with the original DCGAN loss, and the classifier is trained with the classification loss. The generator is trained with all three loss functions because it is the most important neural network in our method. In this subsection, we introduce the individual loss functions.

### 4.2.1 Original Loss

The original GAN loss function is shown in Equation 1. The discriminator is trained to maximize it, and the generator is trained to minimize it. It represents the training philosophy of GANs (i.e., while the generator attempts to deceive the discriminator, it can significantly improve its synthesis capability). We adopt the original loss definitions and denote them as $\mathcal{L}_{orig}^D$ and $\mathcal{L}_{orig}^G$ in our method. They are to train the discriminator and the generator, respectively.

### 4.2.2 Information Loss

To define the information loss, we extract features immediately before the sigmoid activation of the discriminator network (i.e., the flattened tensor marked in yellow in Figure 2). From these features, the discriminator decides whether the inputs are real or synthetic. Thus, it is reasonable to say that the extracted features contain key characteristics of the input samples. In general, extracted features become very high-dimensional vectors after flattening. We use bold font **f** to denote these vectors. The simplest form of information loss is as follows:

$$\mathcal{L}_{mean} = \|\mathbb{E}[\mathbf{f}_x]_{x \sim p_{data}(x)} - \mathbb{E}[\mathbf{f}_{G(z)}]_{z \sim p(z)}\|_2, \quad (2)$$

where **f** stands for features (i.e., high-dimensional vectors) extracted from the last layer of the discriminator and $\mathbb{E}[\mathbf{f}]$ means the average feature (i.e., the centroid of the vectors) of all records in the dataset. Note that we use the L-2 norm (Euclidean norm) to measure the discrepancy between two mean features (vectors).

Thus, $\mathcal{L}_{mean}$ is to compare the first-order statistics (i.e., mean) of the features of real and synthetic records. We also use the second-order statistics (i.e., standard deviation) as follows:

$$\mathcal{L}_{sd} = \|\mathbb{SD}[\mathbf{f}_x]_{x \sim p_{data}(x)} - \mathbb{SD}[\mathbf{f}_{G(z)}]_{z \sim p(z)}\|_2, \quad (3)$$

where $\mathbb{SD}[\cdot]$ represents the standard deviation of features.

$\mathcal{L}_{mean} = 0$ and $\mathcal{L}_{sd} = 0$ mean that real and synthetic records have the statistically same features from the perspective of the discriminator (recall that the extracted features are from the last layer of the discriminator). This further implies that the discriminator may not be able to distinguish them.

The quality of the synthesis process should be controllable. With unreliable partners, you may not want to share a synthetic table that is very similar to the original table. You may want to intentionally generate a low-quality table in that case. For this purpose, we design a loss to train the generator as follows:

$$\mathcal{L}_{info}^G = \max(0, \mathcal{L}_{mean} - \delta_{mean}) + \max(0, \mathcal{L}_{sd} - \delta_{sd}), \quad (4)$$

where $\max(\cdot)$ is used to implement the hinge-loss that does not incur any loss until a predetermined quality degradation threshold

The information loss $\mathcal{L}_{info}^G$ provides zero loss as long as $\mathcal{L}_{mean}$ (resp. $\mathcal{L}_{sd}$) is smaller than a threshold $\delta_{mean}$ (resp. $\delta_{sd}$). Thus, $\delta_{mean}$ and $\delta_{sd}$ are two parameters for controlling the level of privacy. If these parameters are small, then the privacy level will be lower, and the synthetic table will be similar to the original table.

### 4.2.3 Classification Loss

We found that values occasionally do not match with labels in synthetic records, as stated in Section 4.1.3. To avoid this situation, we design an additional loss function called *classification loss* as

$$\begin{aligned}\mathcal{L}_{class}^C &= \mathbb{E}[|\ell(x) - C(remove(x))|]_{x \sim p_{data}(x)}, \\ \mathcal{L}_{class}^G &= \mathbb{E}[|\ell(G(z)) - C(remove(G(z)))|]_{z \sim p(z)},\end{aligned} \quad (5)$$

where $\ell(\cdot)$ is a function that returns the label attribute value of an input record, $remove(\cdot)$ is to remove the label attribute of an

---

**Input:** real records: $\{x_1, x_2, \cdots\} \sim p(x)$
**Output:** a generative model $G$
1   $G \leftarrow$ a generative neural network
2   $D \leftarrow$ a discriminator neural network
3   $C \leftarrow$ a classifier neural network
    /* Initializing to zero vectors         */
4   $\mathbf{f}_{mean}^X \leftarrow \mathbf{0}; \mathbf{f}_{sd}^X \leftarrow \mathbf{0}; \mathbf{f}_{mean}^Z \leftarrow \mathbf{0}; \mathbf{f}_{sd}^Z \leftarrow \mathbf{0}$
5   **while** *until convergence of loss values* **do**
6      Create a mini-batch of real records $X_{mini} = \{x_1, \cdots, x_n\}$
7      Create a mini-batch of latent vector inputs for $G$ $Z_{mini} = \{z_1, \cdots, z_n\}$
8      Perform the SGD update of the discriminator $D$ with $\mathcal{L}_{orig}^D$
9      Perform the SGD update of the classifier $C$ with $\mathcal{L}_{class}^C$
      /* Moving average update of the mean and standard deviation of features */
10     $\mathbf{f}_{mean}^X = w \times \mathbf{f}_{mean}^X + (1-w) \times \mathbb{E}[\mathbf{f}_x]_{x \in X_{mini}}$
11     $\mathbf{f}_{sd}^X = w \times \mathbf{f}_{sd}^X + (1-w) \times \mathbb{SD}[\mathbf{f}_x]_{x \in X_{mini}}$
12     $\mathbf{f}_{mean}^Z = w \times \mathbf{f}_{mean}^Z + (1-w) \times \mathbb{E}[\mathbf{f}_{G(z)}]_{z \in Z_{mini}}$
13     $\mathbf{f}_{sd}^Z = w \times \mathbf{f}_{sd}^Z + (1-w) \times \mathbb{SD}[\mathbf{f}_{G(z)}]_{z \in Z_{mini}}$
14     Perform the SGD update of the generator $G$ with $\mathcal{L}_{orig}^G + \mathcal{L}_{info}^G + \mathcal{L}_{class}^G$
15   **end**
16   **return** $G$

**Algorithm 2:** Training algorithm of table-GAN.

---

input record, and $C(\cdot)$ is a label predicted by the classifier neural network. Thus, this loss is to measure the discrepancy between the label of a generated record and the label predicted by the classifier for that record.

If there are multiple labels, we can extend the classifier neural network to perform a multi-task learning, where the classifier has multiple different final sigmoid activations that share many intermediate layers. Each sigmoid activation is trained to predict a label based on the shared intermediate layers.

We also found that synthetic records occasionally do not recall all values in the original table even after setting $\delta_{mean} = 0$ and $\delta_{sd} = 0$ (i.e., the lowest privacy level and the highest quality in synthetic records). Surprisingly, we discovered in our experiments that the proposed classification loss is able to somehow address this problem in many cases. This is one additional advantage of using the classification loss.

### 4.3 Training Algorithm

In this subsection, we will describe the training algorithm of the proposed table-GAN. In practice, we cannot train neural networks after loading all records simultaneously due to GPU memory limitations. Thus, deep learning algorithms that work with large datasets use the stochastic gradient descent (SGD) update based on mini-batches. We also adopt this approach for better scalability.

However, one problem in the mini-batch training is that we cannot directly calculate the global mean and standard deviation of real and synthetic samples' features (recall that $\mathcal{L}_{mean}$ and $\mathcal{L}_{sd}$ require them). Thus, we use the exponentially weighted moving average to approximate them (lines 10∼13 of Algorithm 2). For instance, $\mathbb{E}[\mathbf{f}_x]_{x \in X_{mini}}$ (resp. $\mathbb{E}[\mathbf{f}_{G(z)}]_{z \in Z_{mini}}$) means the mean feature of a real (resp. synthetic) mini-batch. Using them, it calculates the global mean features $\mathbf{f}_{mean}^X$ and $\mathbf{f}_{mean}^Z$. In general, the weight $w$ should be close to 1 in the moving average calculation to have the stable global mean and standard deviation of features (we used $w = 0.99$).

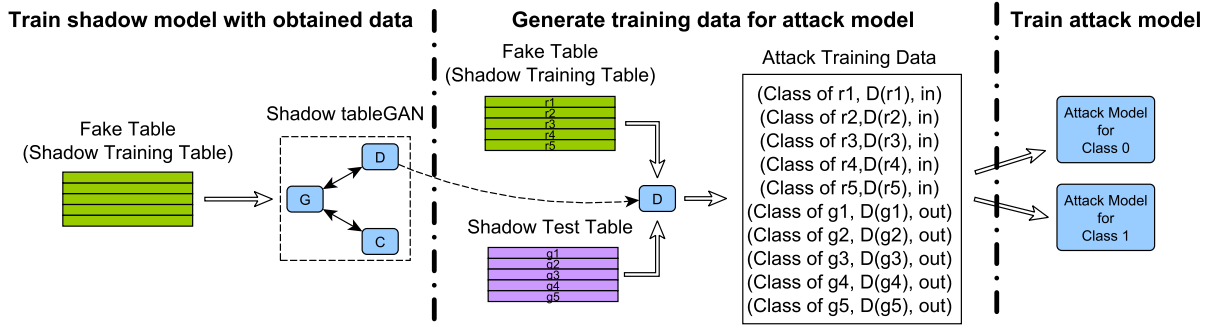

Figure 3: The overall procedure of the customized membership attack method. Its key step is to create a shadow table-GAN, a replica of the target table-GAN that the attacker wants to attack.

The training sequence in an epoch is i) training the discriminator with $\mathcal{L}_{orig}^{D}$ (line 8), ii) training the classifier with $\mathcal{L}_{class}^{C}$ (line 9), and iii) training the generator with $\mathcal{L}_{orig}^{G} + \mathcal{L}_{info}^{G} + \mathcal{L}_{class}^{G}$ (line 14).

Calculating the theoretical complexity for deep learning algorithms is rather cumbersome and meaningless because its training involves many complicated neural network operators (that can be accelerated by GPUs). However, our algorithm requires at most 20 minutes of training time in our experiments.

After being trained, how to generate synthetic records is simple. As shown in Figure 2, the input to the generator is a latent vector $z$. We first randomly sample $z$ in the unit hypercube space and input it to the generator. Its output is one synthetic record. Thus, the generation process is lightweight compared to the training process.

## 4.4 Scalability Issue

There are several standard methods to increase the scalability of deep learning algorithms. Tensorflow supports distributed learning by default and without large efforts, and we can extend to a distributed synthesis method. DownpourSGD, ADMM, EASGD, and GoSGD (all of which are well summarized in [35]) are other famous general distributed learning algorithms. Extensions based on these approaches are also straightforward. An additional advantage of using these approaches is that they perform *ranged-based* search rather than *point-based* search during SGD updates. For the difference between range-based and point-based search, refer to [13].

Another approach to increase the scalability is to i) split a table into several smaller chunks, ii) train a table-GAN with each chunk independently, and iii) generate records with each trained table-GAN and merge them into a synthetic table. Its runtime linearly decreases w.r.t. the number of chunks in this approach. We used this method to synthesize large tables in our experiments.

## 4.5 Membership Attack for table-GAN

To attack the proposed table-GAN, we customize the membership attack method presented in [33] that was designed to attack classification models. Our table-GAN has two classifiers: the discriminator network $D$ and the classifier network $C$. We attack the discriminator $D$ rather than the classifier $C$ because it leads to slightly better success probabilities in our preliminary study. Authors of [33] assume that attackers can i) obtain as many outputs as they want from a target model to attack and ii) know the algorithm and architecture of the target model. As shown in Figure 3, the overall attack procedure is as follows:

1. Let $T$ be a target table-GAN that is already trained and an attacker wants to attack. As in [33], black-box access to the generator of $T$ is allowed. Note that access to two other neural networks of $T$ are blocked because they are not related to releasing synthetic tables after being trained.
2. Obtain many tables synthesized by the generator of $T$, denoted as shadow training table in Figure 3.
3. Train many shadow table-GANs using the obtained synthetic tables.
4. For each trained shadow table-GAN,
   (a) Input each shadow training record $r_i$ to the shadow discriminator $D$ and create an attack training sample of (Class of $r_i$, $D(r_i)$, $in$). Recall that $D(r_i)$ means the predicted probability of being real by the discriminator.
   (b) Input each shadow test record $g_i$ to the shadow discriminator to create a sample (Class of $g_i$, $D(g_i)$, $out$). This shadow test table should consist of real records that are not used to train $T$. In our case, we use the test set prepared for the model compatibility test.
5. Merge all generated attack training samples into one set.
6. Train the attack models using the created attack training data. Now it is ready to attack $T$. Note that we use one attack model per class as in [33] and many state-of-the-art classifiers can be used as attack models.

## 5. EXPERIMENTAL ANALYSIS

We describe the experimental environments and results. We have chosen four tables from four domains: payroll, health, personal records, and airline market. The baseline methods are anonymization/perturbation techniques implemented in ARX [3] and sdcMicro [7], and other generative models. Throughout the experiments, our table-GAN shows the best balance between privacy level and model compatibility.

### 5.1 Experimental Environments

#### 5.1.1 Dataset

We use four tables, as summarized in Table 3. The LACity dataset contains records of Los Angeles city government employees (such as salary, department and so on) [5]. The Adult dataset has many personal records (such as nationality, education level, occupation, work hours per week, and so forth) [1]. The Health dataset consists of various information (such as blood test results, questionnaire survey, diabetes, and so on) [4]. The Airline dataset is created by the Bureau of Transportation Statistics (BTS). BTS randomly selects 10% of all tickets sold in the USA and releases these data to the public every quarter [2]. Each dataset has one

ground-truth label that can be used for model compatibility tests, as follows:

- In the LACity dataset, we know the salary information of employees. Thus, regression analysis tests are available. For classification tests, we use the median salary and create the high-salary attribute. If an employee is paid more than the median salary, then its label is 1.

- In the Adult dataset, the work-hour attribute has the information of work hours per week for each individual. We create additional binary labels after checking whether people work longer than the median case. Thus, we perform both classification and regression tests with this dataset.

- In the Health dataset, we have the diabetes attribute, which indicates whether a person has been diagnosed as having diabetes by doctors. Only classification tests are available in this dataset.

- In the Airline dataset, there is an attribute that contains ticket price information. We can perform regression tests with this attribute. We also create binary labels of whether prices are greater than the median price for classification tests.

We generate synthetic tables that have the same number of records as the original table. For each dataset, we also prepared for testing records that are not part of the original table to check the model compatibility. The number of testing records is approximately 20% of the number of records in the original table.

### 5.1.2 Evaluation Method

The evaluation of data anonymization, perturbation and synthesis methods cannot be performed in a simple manner. This evaluation involves several different methods because they should be evaluated in various aspects. We use the following evaluation methods.

- Privacy-related evaluation metrics are as follows:
  - Recall that existing risk evaluation metrics introduced in Section 2.2 require equivalence classes and one-to-one correspondence between real records and modified records. Our table-GAN cannot be evaluated for them. Instead, we use *distance to the closest record* (DCR) which means the Euclidean distance between a record $r$ of an anonymized, perturbed, or synthesized table and the closest record to $r$ in the original table. Note that an anonymized, perturbed, or synthesized record with DCR = 0 leaks real information. We calculate the distance after attribute-wise normalization because each attribute contributes to the distance equally after the normalization. Distance-based metrics are widely used in many data privacy works [25].
  - Membership attacks are performed following the procedure described in Section 4.5. Note that this attack concept cannot be applied to existing anonymization/perturbation methods because it was designed to attack only machine learning algorithms. We test only our table-GAN for this evaluation.

- Data utility-related evaluation metrics are as follows:
  - *Statistical comparison* is to compare statistical similarity between an attribute in the original table and a corresponding attribute in anonymized, perturbed, or synthesized tables. We will compare cumulative distributions for each attribute.
  - *Machine learning score similarity* is to check the model compatibility. After fixing a classification or regression algorithm and its parameter, we train with the original table or the anonymized/perturbed/synthesized table. If the accuracy values of the two cases for unknown testing cases are the same, then we can say that they are compatible. We use F-1[5] for classification tests and mean relative error (MRE) for regression tests.

Note that in the model compatibility test, we exclude the grid search that is a method to find the best parameter setup. Enabling the grid search during the test means that we compare only the best performing parameter setup for an algorithm. Because two models can show similar accuracy by accident after the grid search, we exclude it.

In general, classical anonymization techniques show good performance for the statistical comparison and machine learning score similarity tests because they do not actively change sensitive attributes. We demonstrate that our method significantly outperforms anonymization techniques with respect to DCR by slightly sacrificing the performance in the statistical comparison and machine learning score similarity tests. Surprisingly, our method shows better model compatibility than anonymization techniques in some tests.

### 5.1.3 Baseline Methods

Among existing anonymization techniques, the combination of $k$-anonymity and $t$-closeness is considered to be the most basic method and is already implemented in ARX, a powerful anonymization tool that is widely used in real applications and research. ARX also has $(\epsilon, d)$-differential privacy and $\delta$-disclosure. Thus, we create two baseline methods with ARX: one is the combination of $k$-anonymity and $t$-closeness and the other is the combination of $(\epsilon, d)$-differential privacy and $\delta$-disclosure. $(\epsilon, d)$-differential privacy requires at least one privacy method that defines equivalence classes and we used $\delta$-disclosure for it. Recall that all these methods in ARX do not change sensitive attributes but construct equivalence classes (by altering QIDs) in a way that can meet the requirements of each privacy protection concept. For perturbation, we will use the micro-aggregation (for QIDs) and the post-randomization method (for sensitive attributes) implemented in sdcMicro. Note that sdcMicro perturbs sensitive attributes as well. All these methods have different privacy requirements. Each tool is able to calculate anonymized/perturbed tables that meet the privacy requirements of those selected mechanisms. Their detailed parameters will be described in Section 5.1.5.

Other baseline methods to generate synthetic records are the condensation method [8] and DCGAN.

### 5.1.4 Computing Environments

We implemented the proposed table-GAN based on Tensorflow and trained in a server with an i7 3.4 Ghz CPU and GTX970 GPU. We did not use any expensive components for the experiments. Even in the entry-level server, it took at most 20 minutes to generate synthetic tables, as shown in Table 4. For the model compatibility test, we programmed using the popular scikit-learn machine learning library [6].

### 5.1.5 Parameter Setups

Table-GAN has two parameters $\delta_{mean}$ and $\delta_{sd}$ to control the level of privacy. We consider the following setups: $\delta_{mean} = 0$ and

---
[5]F-1 is the harmonic mean of precision and recall, and it is one of the most widely used metrics to evaluate classification models.

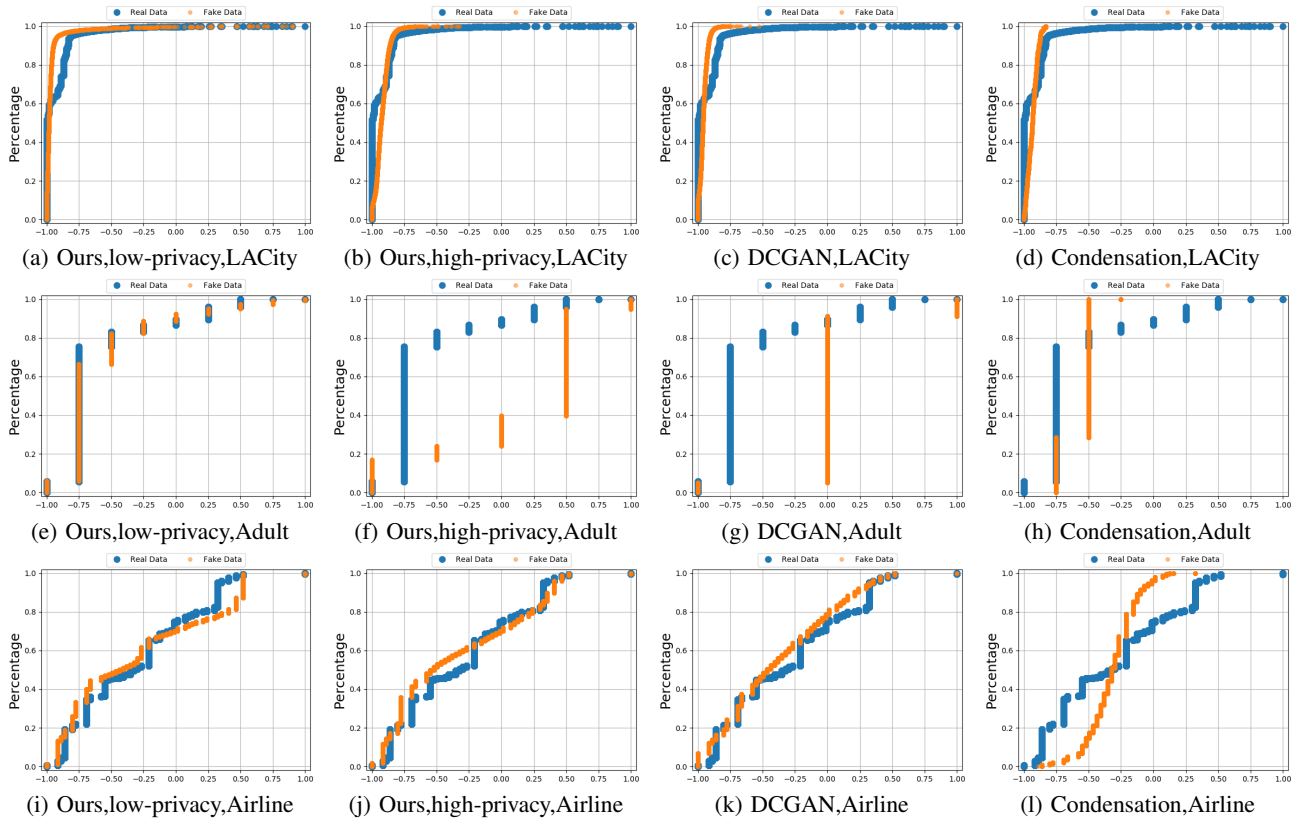

Figure 4: Statistical similarity test: Cumulative distributions of sensitive attributes (base salary, work class, and destination airport ID for each dataset respectively) by the condensation method, DCGAN and table-GAN. X-axes are normalized. Statistics of the original attributes are marked in blue, and synthetic ones are marked in orange. Additional charts for other attributes are in the full version [29].

Table 3: Statistics of datasets

|  | # of Records | # of QIDs | # of Sensitive Attributes | # of Testing Records |
|---|---|---|---|---|
| LACity | 15000 | 2 | 21 | 3000 |
| Adult | 32561 | 5 | 9 | 16281 |
| Health | 9813 | 4 | 28 | 1963 |
| Airline | 1000000 | 2 | 30 | 200000 |

Table 4: Training time of table-GAN in each dataset. We used the multi-chunk parallel approach for Airline to save its training and generating time (see Section 4.4).

|  | LACity | Adult | Health | Airline |
|---|---|---|---|---|
| Training Time of table-GAN | 3.9 mins | 8.16 mins | 1.9 mins | 20.2 mins |

$\delta_{sd} = 0$ as the *low-privacy setting* and $\delta_{mean} = 0.2$ and $\delta_{sd} = 0.2$ as the *high-privacy setting*. With the low-privacy setting, realistic records are generated. By increasing the margins, we generate synthetic tables that are dissimilar to the original table. We train table-GAN for 25 epochs using the Adam optimizer, which is the same as the default DCGAN setting. For the condensation method, we test the condensation group size of 100 and 50.

For $k$-anonymity and $t$-closeness, we tested $k = \{2, 5, 15\}$ and $t = \{0.01, 0.1, 0.5, 0.9\}$. For $(\epsilon, d)$-differential privacy, we used $\epsilon = \{0.01, 0.5, 1, 2, 5\}$, and $d = \{1e-6, 0.001, 0.1\}$; and for $\delta$-disclosure, $\delta = \{1, 2\}$. For sdcMicro, we tested the following parameter setups: $pd = \{0.01, 0.5, 1\}$ and $\alpha = \{0.01, 0.5, 1\}$.

Among these many configurations, we chose the one that leads to the best balance between privacy and model compatibility after testing all parameter setups. In the following subsections, we show the results of the best balance cases for ARX and sdcMicro.

### 5.2 Evaluation Results for Data Utility

In this section, we evaluate the proposed and baseline methods using the statistical and model compatibility tests.

#### 5.2.1 Statistical Comparison

We show cumulative distributions of selected sensitive attributes in Figure 4. We mainly compare the condensation method, DC-GAN and table-GAN after omitting ARX and sdcMicro because they do not significantly change sensitive attributes and show complete matches (i.e., zero-privacy for sensitive attributes) in many cases.

Figures 4 (a-d) are cumulative distributions of the base salary attribute in the LACity dataset. Blue lines are by real values in the original table, and orange lines are by synthetic values. Table-GAN with the low-privacy setting produces a more realistic cumulative distribution and a wider range of values than others. The condensation method and DCGAN do not properly synthesize all values.

Figures 4 (e-h) show four cumulative distributions of the work class attribute in the Adult dataset. Table-GAN with the low-privacy setting shows very good synthesis quality. In Figures 4 (i-l), both of table-GAN and DCGAN successfully reproduce the entire range of values for the destination airport ID attribute in the Airline dataset whereas the condensation method does not show reliable performance. Other attributes have the same pattern in their cumulative distributions (see the full version [29]).

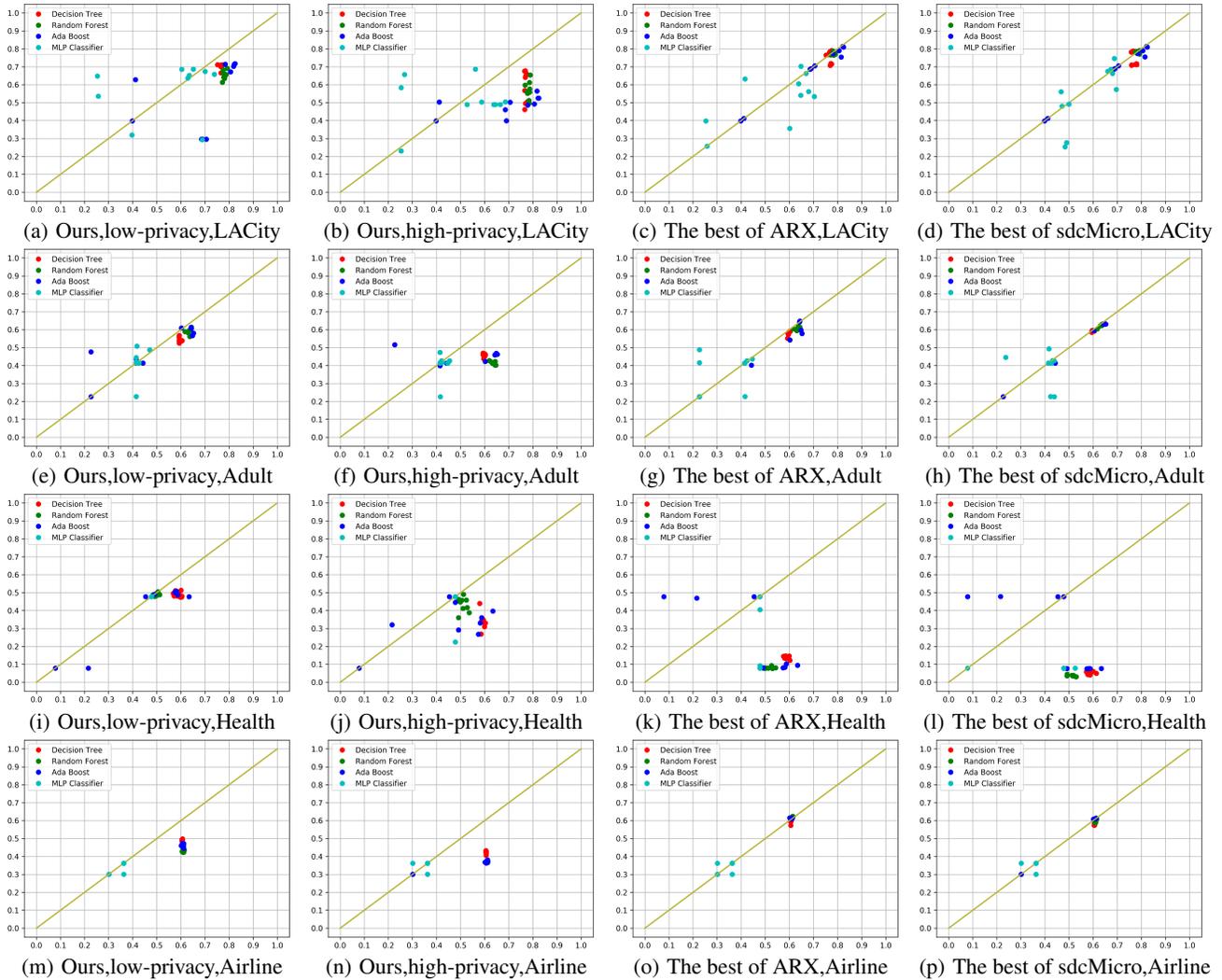

Figure 5: Model compatibility test: Classification score similarity of ARX, sdcMicro, and table-GAN. We remove the condensation method and DCGAN that did not show reliable performance for space limitations. We plot $(x, y)$ after fixing a classification algorithm and its parameter, where $x$ is the F-1 score of the algorithm trained with the original table and $y$ is the F-1 score of the algorithm trained with an anonymized/perturbed/synthesized table. We test 4 algorithms and 10 parameters for each algorithm. Points on the diagonal line (i.e., $x = y$) mean perfect model compatibility. Only our table-GAN shows reliable model compatibility in all datasets.

To summarize, table-GAN with the low-privacy setting shows very high-quality synthesis performance. In all cases, synthetic tables are statistically similar to the original table. DCGAN performs poorly in many cases because its loss function is not designed for the purpose of table synthesis. Table-GAN with the high-privacy setting performs better than DCGAN. The condensation method shows the worst synthesis performance. Surprisingly, its overall synthesis quality is worse than DCGAN. We think that this is because DCGAN, whose loss functions are defined for image synthesis, is able to capture more statistics of records.

### 5.2.2 Model Compatibility

We perform several classification and regression tasks to check the model compatibility. We perform in-depth analyses and prove that our table-GAN shows the best balance between privacy level and model compatibility. We do not list the results of the condensation method and DCGAN for their poor synthesis performance and space limitations.

#### 5.2.2.1 Classification.

In Figure 5, we plot $(x, y)$ pairs, where $x$ is the F-1 score of the model trained with the original table and $y$ is that of the model trained with an anonymized, perturbed, or synthesized table in each dataset. We check scores for unknown testing records. Recall that we exclude the grid search, and every $x$ and $y$ pair is calculated using the same machine leaning algorithm with the same parameter setup. We used decision tree, random forest, AdaBoost, and multi-layer perception classifiers and 10 parameter setups for each algorithm (i.e., 40 points in total in a plot) — we referred to the scikit-learn web pages to collect recommended parameter setups. The diagonal line represents perfect model compatibility (i.e., $x = y$ and anonymized/perturbed/synthetic tables train machine learning algorithms in the same way as the original table). For ARX, we choose the best configuration that shows the best model compatibility in each dataset.

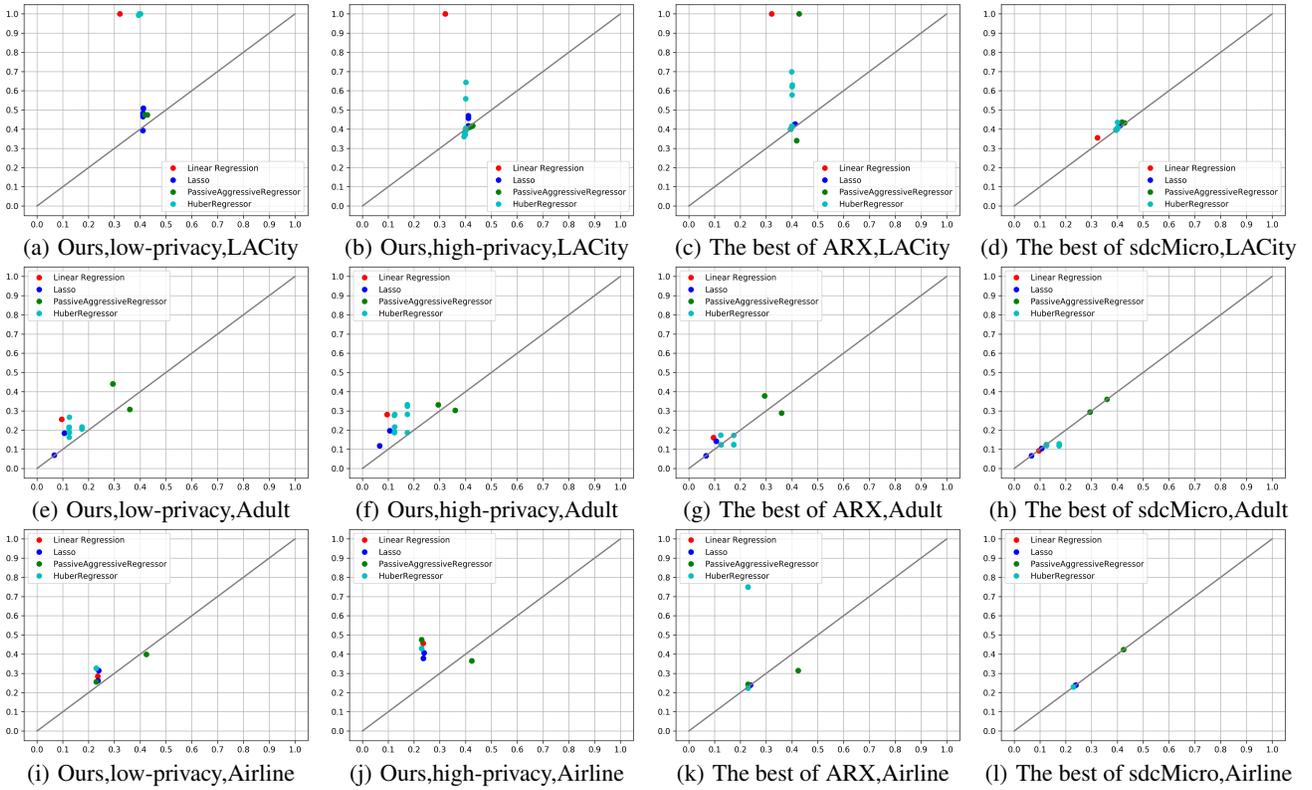

Figure 6: Model compatibility test: Regression score similarity of ARX, sdcMicro, and table-GAN. We remove the condensation method and DCGAN that did not show reliable performance for space limitations. We plot $(x, y)$ after fixing a regression algorithm and its parameter, where $x$ is the mean relative error (MRE) score of the algorithm trained with the original table and $y$ is the MRE score of the algorithm trained with an anonymized/perturbed/synthesized table. We test 4 algorithms and 10 parameters for each algorithm. Points on the diagonal line (i.e., $x = y$) mean perfect model compatibility.

Figures 5 (a-d) show the F-1 score similarity in the LACity dataset. The tables anonymized by ARX (5-anonymity and 0.01-closeness) and sdcMicro show the best model compatibility in (c) and (d), which is very clear because their modifications to any sensitive attributes are very limited, as shown in the previous DCR tests. ARX with $(\epsilon, d)$-differential privacy and $\delta$-disclosure does not show as good model compatibility as them and we removed it for space limitations. QIDs are also important features in this dataset. Thus, they do not show perfect model compatibility due to the modified QID values[6]. Table-GAN with the low-privacy setting in (a) shows the second-best model compatibility with very small differences.

Figures 5 (e-h) show the test results from the Adult dataset. Surprisingly, our table-GAN with the low-privacy setting in (e) shows a model compatibility that is slightly worse than the best ARX or sdcMicro cases. In many cases, points by both methods are around the diagonal line. For the Health dataset in Figures 5 (i-l), our table-GAN shows better model compatibility than all other baseline methods. Only our table-GAN shows practically meaningful model compatibility in this dataset.

Classification scores in the Airline dataset are in Figures 5 (m-p). ARX and sdcMicro show very good model compatibility. Table-GAN with the low-privacy setting is slightly worse than them. However, its model compatibility is still acceptable.

---

[6] We applied the label encoding algorithm implemented in scikit-learn if modified QIDs are not numerical values.

#### 5.2.2.2 Regression.

In Figure 6, we show the results of the regression model compatibility tests. We follow the same plotting method that shows $(x, y)$ scores. We use mean relative error (MRE) as a base metric to evaluate regression models. Points on the diagonal line means perfect model compatibility. We use the following four regression algorithms and 10 parameter setups for each algorithm: linear regression, Lasso regression, passive aggressive regression, Hurber regression. Because the Health dataset has only binary labels, we cannot perform regression tests.

In almost all datasets, table-GAN, ARX and sdcMicro show very good model compatibility. In general, sdcMicro shows better model compatibility than others because its modifications on data is very limited (i.e., low privacy). Our table-GAN shows better model compatibility than ARX.

### 5.3 Evaluation Results for Privacy

We evaluate the proposed and baseline methods for privacy using the distance-based metric and membership attack. In particular, the distance-based evaluation is one of the most basic metrics to check privacy level [25].

#### 5.3.1 Distance to the Closest Record

Let $r$ be a record in the original table. In an anonymized or perturbed table, a record $r'$ that is modified from $r$ always exists. Therefore, their relationship is bijective and weak from re-identification attacks in many cases. In synthetic tables, however, there is no such relationship and thus, we instead find the synthetic record closest to $r$ in Euclidean distance.

Table 5: Euclidean distance between a real record to its closest synthetic/anonymized/perturbed record after attribute-wise normalization. Its format is average $\pm$ std. dev. in each cell. Risky cases, where the average distance is too small or the standard deviation is too large, are indicated in red.

|  | table-GAN (low-privacy) | table-GAN (high-privacy) | The best of ARX | The best of sdcMicro | DCGAN |
|---|---|---|---|---|---|
| QIDs + Sensitive attributes ||||||
| LACity | $0.96 \pm 0.22$ | $1.48 \pm 0.3$ | $0.68 \pm 0.52$ | $0.07 \pm 0.17$ | $0.83 \pm 0.31$ |
| Adult | $0.75 \pm 0.19$ | $1.84 \pm 0.23$ | $0.59 \pm 0.17$ | $0.54 \pm 0.12$ | $0.88 \pm 0.24$ |
| Health | $2.53 \pm 0.43$ | $2.75 \pm 0.41$ | $0.61 \pm 0.25$ | $1.23 \pm 0.34$ | $2.85 \pm 0.42$ |
| Airline | $1.21 \pm 0.21$ | $1.23 \pm 0.27$ | $1.46 \pm 0.32$ | $0.98 \pm 0.41$ | $0.86 \pm 0.15$ |
| Only Sensitive attributes ||||||
| LACity | $0.68 \pm 0.18$ | $1.24 \pm 0.17$ | $0 \pm 0$ | $0.05 \pm 0.13$ | $0.54 \pm 0.18$ |
| Adult | $0.45 \pm 0.14$ | $1.25 \pm 0.17$ | $0 \pm 0$ | $0.2 \pm 0.1$ | $0.82 \pm 0.24$ |
| Health | $2.4 \pm 0.38$ | $2.56 \pm 0.39$ | $0 \pm 0$ | $0.22 \pm 0.2$ | $2.68 \pm 0.41$ |
| Airline | $0.96 \pm 0.19$ | $1.08 \pm 0.26$ | $0 \pm 0$ | $0.69 \pm 0.36$ | $0.76 \pm 0.16$ |

Table 6: The evaluation of membership attacks

| Dataset | $\delta_{mean} = \delta_{sd} = 0$ (low-privacy) | | $\delta_{mean} = \delta_{sd} = 0.1$ (mid-privacy) | | $\delta_{mean} = \delta_{sd} = 0.2$ (high-privacy) | |
|---|---|---|---|---|---|---|
|  | F-1 | AUCROC | F-1 | AUCROC | F-1 | AUCROC |
| LACity | 0.59 | 0.64 | 0.49 | 0.6 | 0.4 | 0.46 |
| Adult | 0.51 | 0.49 | 0.41 | 0.5 | 0.19 | 0.5 |
| Health | 0.33 | 0.48 | 0.34 | 0.5 | 0.3 | 0.45 |
| Airline | 0.54 | 0.5 | 0.48 | 0.47 | 0.45 | 0.47 |

Table 5 shows the average and standard deviation of distances of $(r, c)$ pairs, where $r$ is an original record and $c$ is the record closest to $r$ in an anonymized/perturbed/synthesized table. It is preferred that the average distance is large and the standard devision is small. A large standard deviation means, even though its average distance is large, there exist some $(r, c)$ pairs that are very close.

As expected, ARX did not change any sensitive values, and its average distance values are always zeros when considering only sensitive attributes. Table-GAN with the low-privacy setting shows at most tens of times longer average distance values (and thus, substantially lower probabilities of privacy leakage) than ARX and sdcMicro. Our table-GAN shows very stable average and standard deviation values. Moreover, there is no one-to-one relationship between the original and generated tables. In fact, it is almost impossible to re-identity original values from synthetic values (see our generation examples in Table 7 and 8).

### 5.3.2 Membership Attack

We attack our table-GAN with various hinge-loss parameter configurations. For each dataset, two attack models (one per class) are prepared following the procedure outlined in Section 4.5. We also prepare for a balanced set of attack testing records (i.e., 50% of 'in' and 50% of 'out' records) for each dataset — the 'in' records are from the original training table; and the 'out' records are from the model compatibility testing records that are reserved for this purpose and have not been used to create class-based attack models. We use Multilayer Perceptron, DecisionTree, AdaBoost, RandomForest, and SVM classifiers to build attack models and their best parameters are found through the grid search with 10-fold cross validation. We evaluate the attack performance based on F-1 and AUCROC.

In Table 6, we summarize the attack performance (averaged over two classes) — we observe that the performance does not greatly depend on classes. In almost all cases, the low-privacy setting allows some information leakage to the attacker, resulting in the F-1 and AUCROC values of up to 0.64. As two hinge-loss parameters, $\delta_{mean}$ and $\delta_{sd}$, increase, the attack performance decreases. In the Adult dataset, for instance, F-1 is dropped from 0.51 to 0.19, which means unsuccessful membership attacks. In many other cases, the attack performance is decreased by 10% in the high-privacy configuration, compared to that in the low-privacy configuration.

Table 7: Sample records in the original LACity table

| Year | Salary | Q1 | Q2 | Q3 | Dept | Job |
|---|---|---|---|---|---|---|
| 2014 | 70386.48 | 16129.89 | 17829.78 | 17678.24 | 98 | 1230 |
| 2013 | 52450.56 | 11331 | 13859.93 | 11968.32 | 70 | 2214 |
| 2013 | 89303.76 | 20036.32 | 23479.2 | 21153.6 | 70 | 2214 |
| 2013 | 60028.96 | 15793.88 | 18560.38 | 16471.18 | 42 | 3184 |
| 2014 | 64553.13 | 14700 | 17313.1 | 15257.17 | 82 | 1368 |
| 2014 | 65959.92 | 26530.26 | 32978.41 | 25697.5 | 98 | 3181 |

Table 8: Sample records in the synthesized table by table-GAN with the low-privacy setting. For each real record in Table 7, we have chosen the closest synthetic record in Euclidean distance.

| Year | Salary | Q1 | Q2 | Q3 | Dept | Job |
|---|---|---|---|---|---|---|
| 2013 | 72005.93 | 11747.34 | 17186.00 | 19557.64 | 50 | 1451 |
| 2013 | 59747.90 | 4369.88 | 13377.60 | 22311.95 | 73 | 1248 |
| 2013 | 85600.46 | 17993.01 | 25420.13 | 27127.87 | 46 | 2025 |
| 2013 | 65156.87 | 11011.99 | 20201.47 | 23563.72 | 67 | 1887 |
| 2014 | 68638.75 | 9642.26 | 13674.69 | 15680.99 | 51 | 998 |
| 2014 | 73140.91 | 14474.15 | 28872.33 | 30307.91 | 71 | 2279 |

## 5.4 Generation Example

We show generation examples based on the LACity dataset in Table 8. Sample records of the original LACity table are shown in Table 7 after selecting a subset of columns for space reasons.

Our table-GAN (with the low-privacy setting) generates Table 8. Because there is no one-to-one correspondence between Table 7 and Table 8, we select a synthetic record closest to each real record of Table 7 after attribute-wise normalization. As shown, real records have very different values from the closest synthetic record. It is almost impossible to identify original records in the synthetic table.

## 6. CONCLUSION

We proposed a method called table-GAN to synthesize tables. Our method shows as good model compatibility as anonymization techniques that do not change sensitive attributes. To our knowledge, our method is the first attempt to synthesize general relational databases using deep learning techniques.

We performed experiments using four real-world datasets with millions of records and tens of attributes. The runtime to synthesize such large databases is approximately 20 minutes with our entry-level servers due to the advancements in GPU technology. In the statistical, distance to the closest record, and model compatibility tests, table-GAN exhibits the best trade-off between privacy level and model compatibility.

In the future, we plan to extend our method to other data types such as strings and further improve the generation quality. We hypothesize that data synthesis based on well-designed generative models can lead to perfect model compatibility.

## 7. ACKNOWLEDGMENTS

This work was supported by the National Research Council of Science & Technology (NST) grant by the Korea government (MSIP) [No. CRC-15-05-ETRI]. We thank Dr. Sungjin Ahn in Rutgers University for his comments during the early phase of the research. We also thank anonymous reviewers for sharing their ideas to better evaluate the privacy risk in our method.

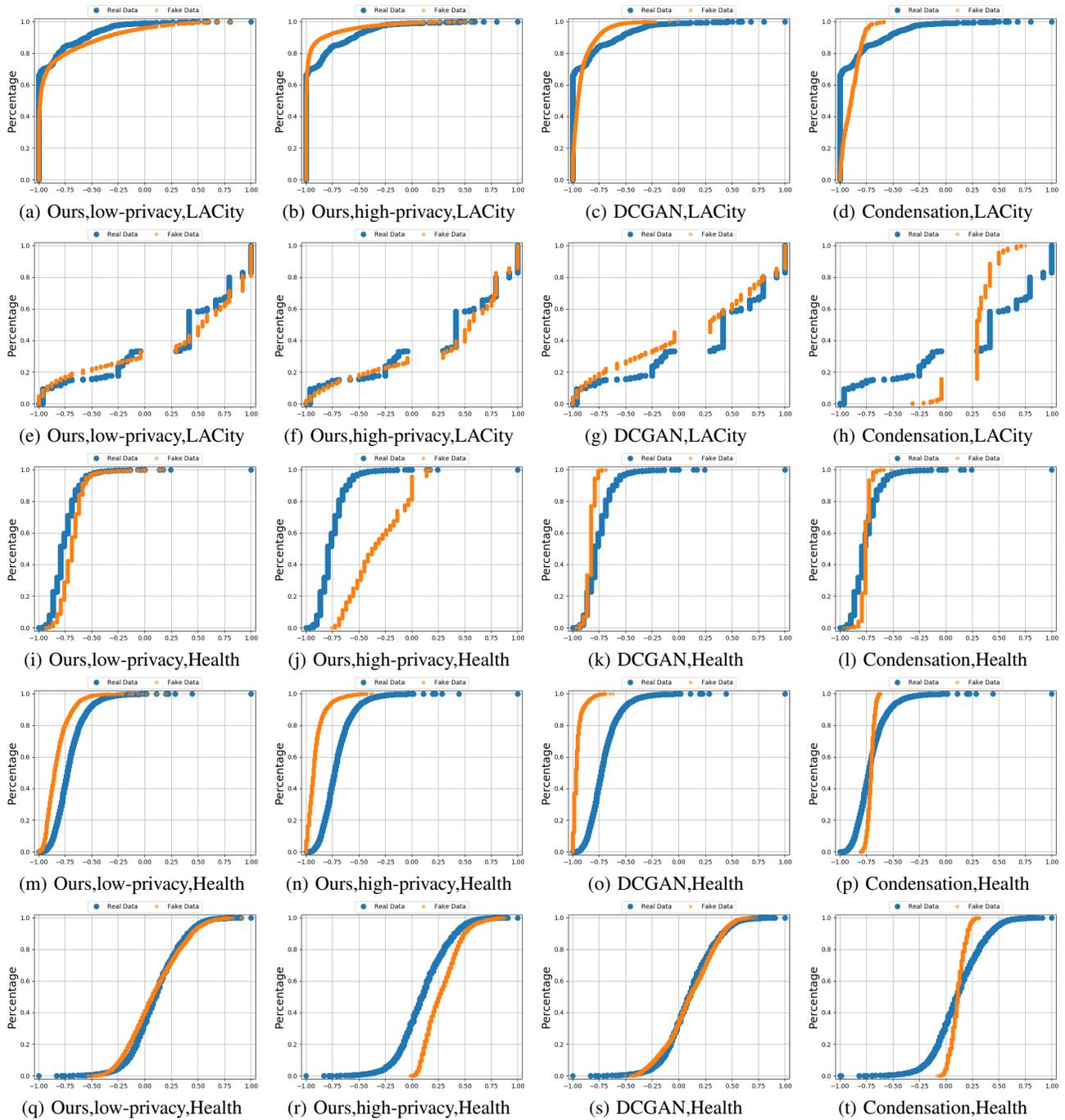

Figure 7: **Cumulative distributions of some sensitive attributes in the LACity and Health dataset**

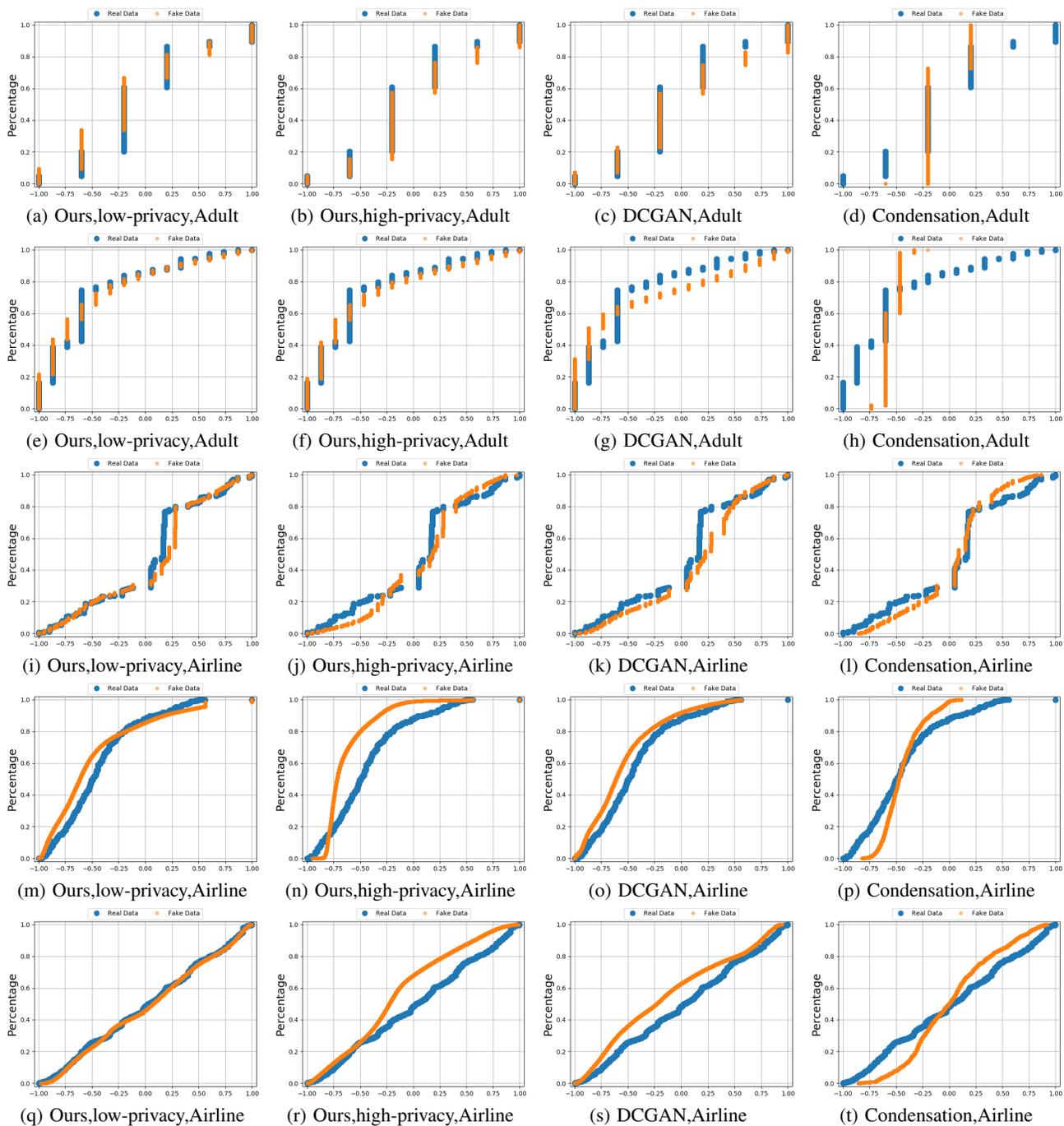

Figure 8: Cumulative distributions of some sensitive attributes in the Adult and Airline dataset

# APPENDIX
## A. STATISTICAL COMPARISON

We show more cumulative distribution comparisons in Figures 7 and 8. Our table-GAN with the low-privacy setting shows very good generation performance in all datasets. The condensation method and DCGAN occasionally shows acceptable generations.